\documentclass[12pt,preprint]{aastex}

\slugcomment{Scheduled for publication in ApJ, 2005 Feb 10}
\shorttitle{Sub-mm images of $\eta$ Corvi}
\shortauthors{Wyatt et al.}

\received{2004 May 27}
\begin{document}

\title{Sub-millimeter images of a dusty Kuiper belt around
$\eta$ Corvi}

\author{M. C. Wyatt\altaffilmark{1},
        J. S. Greaves\altaffilmark{2},
        W. R. F. Dent\altaffilmark{1},
        I. M. Coulson\altaffilmark{3}}

\altaffiltext{1}{UK Astronomy Technology Centre, Royal Observatory,
  Edinburgh EH9 3HJ, UK}
\altaffiltext{2}{School of Physics and Astronomy, University of St Andrews, 
  North Haugh, St Andrews KY16 9SS, UK}
\altaffiltext{3}{Joint Astronomy Centre, 660 N. A'ohoku Place, Hilo,
  HI96720, USA}

\begin{abstract}
We present sub-millimeter and mid-infrared images of the circumstellar
disk around the nearby F2V star $\eta$ Corvi.
The disk is resolved at 850 $\mu$m with a size of $\sim 100$ AU.
At 450 $\mu$m the emission is found to be extended at all
position angles, with significant elongation along a position
angle of $130 \pm 10^\circ$;
at the highest resolution ($9\farcs 3$) this emission is resolved
into two peaks which are to within the uncertainties offset
symmetrically from the star at 100 AU projected separation.
Modeling the appearance of emission from a narrow ring in the
sub-mm images shows the observed structure cannot be caused by an
edge-on or face-on axisymmetric ring;
the observations are consistent with a ring of radius $150 \pm 20$
AU seen at $45 \pm 25^\circ$ inclination.
More face-on orientations are possible if the dust distribution includes
two clumps similar to Vega;
we show how such a clumpy structure could arise from the migration
over 25 Myr of a Neptune mass planet from 80-105 AU.
The inner 100 AU of the system appears relatively empty
of sub-mm emitting dust, indicating that this region may have been
cleared by the formation of planets, but the disk emission
spectrum shows that IRAS detected an additional hot component with a
characteristic temperature of $370 \pm 60$ K (implying a distance
of 1-2 AU).
At 11.9 $\mu$m we found the emission to be unresolved with no
background sources which could be contaminating the fluxes
measured by IRAS.
The age of this star is estimated to be $\sim 1$ Gyr.
It is very unusual for such an old main sequence star to exhibit
significant mid-IR emission.
The proximity of this source makes it a perfect candidate
for further study from optical to mm wavelengths
to determine the distribution of its dust.
\end{abstract}


\section{Introduction}
\label{s:intro}
The infrared satellite IRAS found that some 15\% of nearby main sequence
stars exhibit infrared emission in excess of that expected from the stellar 
photosphere alone (e.g., Aumann et al. 1984; Backman \& Paresce 1993).
This excess emission is thought to come from dust which is heated by the star
and its temperature ($<120$ K) implies that for
the majority of stars this dust is in regions analogous to the Kuiper
belt in the Solar System ($>30$ AU; Wyatt et al. 2004).
This analogy is reinforced by detailed analysis of these systems which show
that the dust must be continually replenished from a reservoir of larger
(km-sized) planetesimals (Wyatt \& Dent 2002).
Very few stars have been seen to exhibit warm emission from dust at
$<20$ AU from the star;
those that do tend to be young with age $<400$ Myr (Laureijs et al.
2002), an observation which may indicate that these inner regions have
been cleared by the formation of planets (e.g., Williams et al. 2004).
Further indirect evidence for the presence of planets in these inner
regions comes from the clumpy and asymmetric structures of the disks
(Holland et al. 1998; Greaves et al. 1998);
modeling has shown that these structures may be caused by the effect
of an unseen planet's gravity on the disk (Wyatt et al. 1999; Ozernoy et
al. 2000; Wyatt 2003). 
Thus these debris disks play a fundamental role in increasing our
understanding of the outcome of planet formation.

While several hundred stars are now known to exhibit infrared excesses,
the next fundamental step is to image this emission.
Such images are important for a number of reasons:
Imaging confirms that the IRAS excess is associated with star, since
the large IRAS beam size means that some excess sources are confused
with background objects (Lisse et al. 2002; Jayawardhana et al. 2002;
Sheret, Dent, \& Wyatt 2004).
It also gives a direct measure of the radial location of dust from which
conclusions can be drawn about the size and emptiness of the inner
regions, as well as about the size and composition of the dust (Wyatt \&
Dent 2002; Sheret et al. 2004).
The detection of asymmetric structures in the images can also be used to
set constraints on an unseen planetary system in the inner region (Wyatt 2003).
Once this type of detailed information has been accumulated for a
sufficient number of disks it will be possible to consider their
evolutionary sequence and the diversity of possible outcomes of planet
formation around different types of star.

Until recently only six disks had been imaged
at wavelengths ranging from the optical (Clampin et al. 2003) through the
mid-IR (Telesco et al. 2000), to the sub-mm (Holland et al. 1998; Greaves et al.
1998; Holland et al. 2003), and the mm (Koerner et al. 2001; Wilner et al. 2001).
Not surprisingly these were also the six brightest candidates found by
IRAS.
Five of these disks are found around bright A stars, and one around the
nearby K star $\epsilon$ Eri.
The challenge with imaging of debris disks is that they are faint;
sub-mm imaging of the brightest disks required several nights of
telescope time.
Given that fluxes and size scales have large uncertainties based on the IRAS
fluxes it is prohibitive to study all IRAS excess sources in the same
detail.
Recent observational campaigns have had success at resolving disks by
focussing on young stars (Williams et al. 2004; Kalas, Liu \& Matthews
2004);
the disks of stars younger than a few 100 Myr might be expected to be
brighter than those of older stars, since there is evidence that disk
mass decreases with age (Habing et al. 2001; Greaves \& Wyatt 2003).
However, a disk was recently resolved in sub-millimeter imaging around a
star which is older than the Sun (Greaves et al. 2004).
Since the three newly resolved disks are around G and M stars, these
new images are filling in our understanding of how disk parameters
are affected by spectral type and age.

Here we report on the discovery of a resolved disk around the nearby
star $\eta$ Corvi (HD109085), located at $18.2 \pm 0.2$ pc from the Sun.
The star is of spectral type F2V (Mora et al. 2001) and has an age
of $\sim 1$ Gyr (see \S \ref{s:concl}).
Its disk is the first to be resolved around a main sequence F star.
Previous surveys of the IRAS database found this star to exhibit an
infrared excess (Stencel \& Backman 1991; Mannings \& Barlow 1998),
and a recent sub-mm photometric survey concluded that this disk should
be large enough to be resolved with sub-mm imaging (Sheret et al. 2004).
Sheret et al. also noted that there is a large mid-IR excess in this
system which is expected to be at a different radial location to the
sub-mm emission.

The structure of this paper is as follows.
Sub-mm imaging observations of this system are described in \S
\ref{s:obs} and the results presented in \S \ref{s:res}.
Mid-IR observations of this system are also described in \S \ref{s:obs}
and \S \ref{s:res}.
Modeling of the observations and of the disk's spectral energy distribution
are presented in \S \ref{s:disc} and a discussion of the implications for
the nature of this disk are given in \S \ref{s:concl}.

\section{Observations}
\label{s:obs}

\subsection{Sub-mm}
\label{ss:submm}
The sub-mm observations of $\eta$ Corvi were made using the Submillimetre
Common-User Bolometer Array, SCUBA at the James Clerk Maxwell telescope (JCMT).
SCUBA images at wavelengths of 850 and 450 $\mu$m simultaneously
onto arrays containing 91 and 37 bolometers respectively covering
a total field of view of $\sim 2\farcm 3$ diameter (Holland et al.
1999).
Each bolometer has diffraction-limited resolution on the sky
with beam sizes of $14\arcsec$ and $8\farcs 1$ at 850 and 450 $\mu$m
respectively.
We observed $\eta$ Corvi ($12^h32^m04\fs 23$, $-16^\circ 11'45\farcs 6$)
for a total of 13.5 hours over 10 nights (6.4 hours in 2003 January 5-9 and
7.1 hours in 2003 March 13-16 and 21) at an average airmass of 1.4.
All nights had excellent weather conditions with zenith opacities in
the range 0.5-0.9 at 450 $\mu$m and 0.12-0.20 at 850 $\mu$m.
Since the arrays instantaneously undersample the sky, a fully sampled
image was achieved using the standard technique of jiggle-mapping. 
In this method the final map is made up of a co-add of 2 second
jiggle observations in which the arrays are pointed at different
offsets from the star using the secondary mirror (Holland et al. 1999).

Given the small scale of the predicted structure,
the pointing was checked thoroughly before and after each 20-40 minute
observation on pointing sources 1213-179 or 3C279.
The pointing accuracy was found to be $\sim 1\farcs 7$, slightly better
than the $\pm 2\arcsec$ errors quoted for the JCMT.
\footnote{The accuracy quoted here is the difference in pointing offsets
for the pointings before and after the observations.
This was corrected in the reduction by setting the pointing midway between
the two pointing observations (i.e., implying that pointing accuracy is
$\pm 0\farcs 8$).
In all analysis the tracking error due to the F2 correction
(http://www.jach.hawaii.edu/JACpublic/JCMT/Facility\_description/Pointing/tracking\_fault.html)
was completely removed in the data reduction;
the maximum correction applied during any one observation was $2\farcs 4$.}
The data for each night were calibrated using Mars when available,
or one of the standard secondary JCMT calibrators.
Calibration uncertainties were estimated from the night to night
gain variation as $\pm 9\%$ at 850 $\mu$m and $\pm 13\%$ at 450 $\mu$m.
Point source calibration observations were modeled to show that the PSF of
the SCUBA observations could be well fit by Gaussians with FWHM of
$14\arcsec$ and $8\farcs 1$ at 850 and 450 $\mu$m respectively.

The first part of the data reduction was accomplished using the
\textit{SURF} package (Jenness \& Lightfoot 1998).
The data were corrected for atmospheric extinction using 
850 and 450 $\mu$m sky opacities derived
from the CSO's 225GHz (January) and 350 $\mu$m
(March) dipper measurements (Weferling 2004).
The CSO dip measurements were taken at 10 minute intervals
and their relation to sub-mm opacities is
well understood (Archibald et al. 2002).
Each 20-40 minute observation was reduced assuming an opacity
derived from a polynomial fit to the long term trend of the opacity
variation on each night.
The accuracy of this method was confirmed using observations
of the 850 $\mu$m opacity measured less frequently at the JCMT
using the skydip method (Archibald et al. 2002).

The standard observational technique of sky chopping (at $\sim 8$ Hz)
was used to remove the dominant sky background during each observation;
the chop throw employed was $60\arcsec$ in January and $120\arcsec$
in March.
The telescope primary was nodded (every 16 seconds) to place the source
alternately in both chop beams and thus cancel slower varying sky gradients.
Short timescale spatial and temporal variations in sky emissivity
("sky-noise") were further removed using the \textit{SURF} routine
REMSKY which subtracts the mean level of bolometers not on source during
each 2 second jiggle observation.
However, even employing these sky subtraction methods a low-level
residual background signal remained in each 20-40 minute observation
at a level of $\pm 3$ and $\pm 19$ mJy/beam at 850 and
450 $\mu$m respectively.
This DC-offset is believed to be caused by residual beam imbalance
(not cancelled by nodding) due to a time-dependent spillover signal,
most likely from beyond the edge of the primary as the secondary chops
and jiggles during a 20-40 minute observation.
This residual sky level was removed by subtracting the mean
surface brightness $>30\arcsec$ from the star from each observation.
Since the DC level is constant across the array the subtraction does not
result in higher resolution structure being introduced into the final map.

The final part of the reduction was accomplished using custom
routines.
The data for observations on all nights were rebinned,
with the measurements from different bolometers (on different nights)
weighted according to their noise;
this noise level was estimated from the standard deviation of flux
measurements made by each bolometer in an observation.
\footnote{While the jiggle pattern may mean that a bolometer moves on
and off source during an observation, this does not affect the noise
estimate since the source signal is at least ten times fainter the
noise in a single 2 second exposure.}
Bolometers with above average noise on any given night were not
included in the final map, and anomalous signals were clipped above
the 3 $\sigma$ level.
The resulting maps do not differ significantly to those produced using
\textit{SURF}'s rebin, however the routines also return a map of the
uncertainty in the flux measurements in the map.

\subsection{Mid-IR}
\label{ss:midir}
The mid-IR observations were performed using the Thermal Infrared
MultiMode Instrument TIMMI2 at the ESO 3.6m telescope at La Silla.
We imaged $\eta$ Corvi for 840 seconds in 2003 November 19 at
an airmass of 1.7.
The observations used the instrument's 11.9 $\mu$m
filter and a $0\farcs 2$ pixel scale was chosen giving a total
field of view of 64\arcsec x 48\arcsec on the 320x240 pixel
array.
The sky background was removed by chopping North-South with $10\arcsec$
amplitude and the telescope emission and sky residuals were
subtracted by nodding East-West, again with $10\arcsec$ amplitude.
This observing strategy means that a straight co-add of the data
results in an image which contains two positive and two negative
images of any given source.
The image was calibrated using observations of the standard star
HD71701 taken immediately before the science observation at an
airmass of 1.5;
the flux of the calibrator in this waveband, 6312 mJy, was derived
from the template for the emission spectrum of this star given
in Cohen et al. (1999).
A conservative estimate of the calibration accuracy was
determined from the gain variation throughout the night to be
$\pm 16$\%.
The PSF of point sources on this night had FWHM of between
$1\farcs0$ and $1\farcs2$.

\section{Results}
\label{s:res}

\subsection{Sub-mm}
\label{s:submmresults}
\begin{figure}
  \begin{center}
    \begin{tabular}{c}
        \epsscale{1.0} \plotone{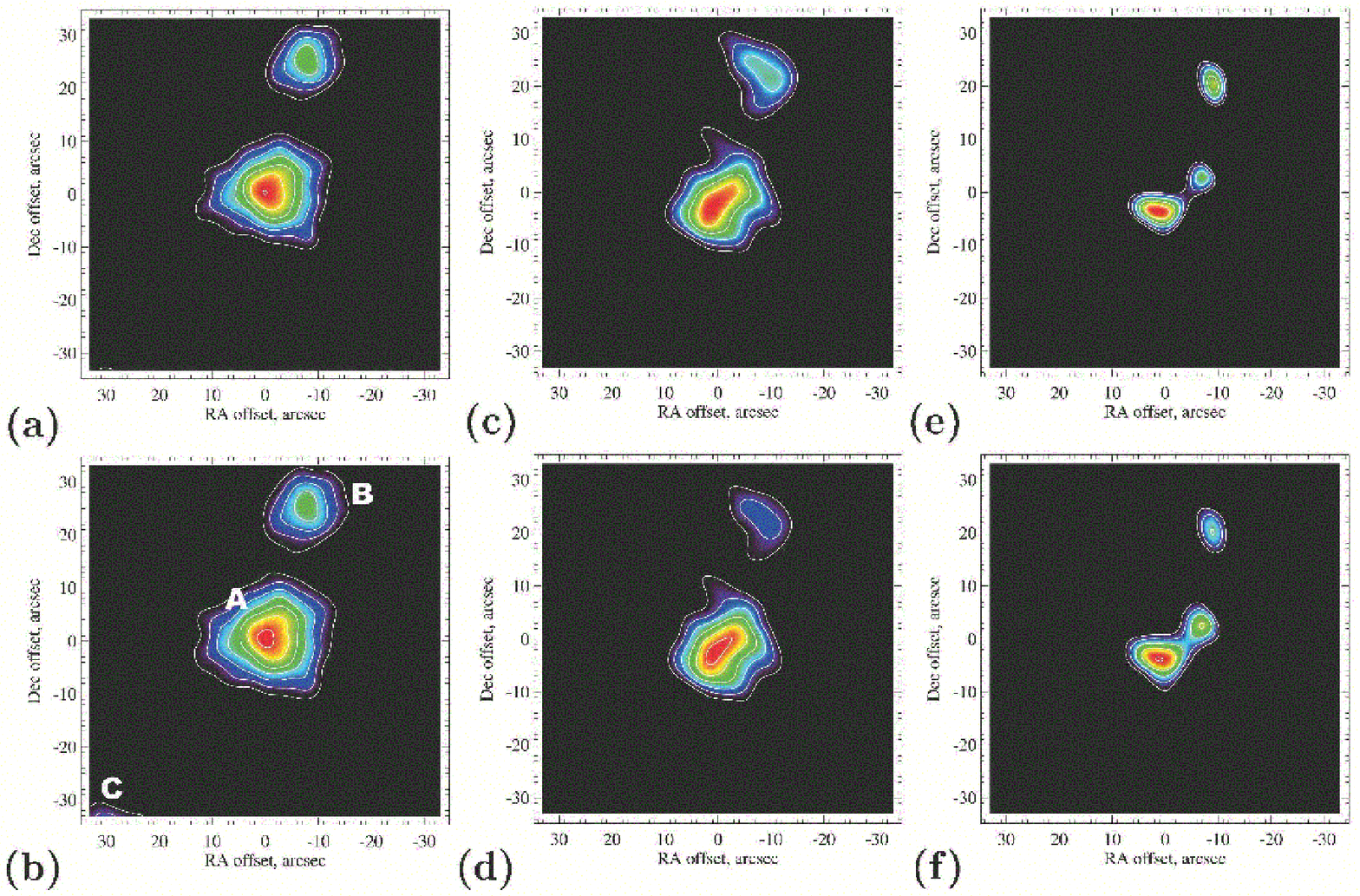}
    \end{tabular}
  \end{center}  
  \caption{SCUBA observations of $\eta$ Corvi at:
  \textbf{(a)} and \textbf{(b)}
  850 $\mu$m with an effective resolution of $15\farcs 8$;
  \textbf{(c)} and \textbf{(d)}
  450 $\mu$m with an effective resolution of $13\farcs 7$;
  and \textbf{(e)} and \textbf{(f)}
  450 $\mu$m with an effective resolution of $9\farcs 5$.
  The top row of images \textbf{(a,c,e)} show the surface brightness
  of the emission, while the bottom row \textbf{(b,d,f)} show the
  signal to noise of that emission.
  All maps are presented with a small amount of
  additional Gaussian smoothing:
  \textbf{(a)} and \textbf{(b)} $3\arcsec$;
  \textbf{(c-f)} $4\arcsec$.
  Contour levels in the surface brightness images are linearly spaced at:
  \textbf{(a)} [16, 20, ...],
  \textbf{(c)} [72, 81, ...],
  and \textbf{(e)} [110, 118, ...] $\mu$Jy arcsec$^{-2}$.
  Contour levels in the signal to noise images are linearly spaced at:
  \textbf{(b)} [3, 4, ...],
  \textbf{(d)} [4, 4.5, ...],
  and \textbf{(f)} [3.2, 3.6, ...].
  The axes show the offsets from the location of the star;
  N is up, E is to the left.
  The structure seen within $15\arcsec$ of the star in \textbf{(e)} and
  \textbf{(f)} is consistent with two clumps of equal brightness both
  offset $5\farcs 5$ from the star (see text for details).
  \label{fig:scuba}}
\end{figure}

The rebinned 850 and 450 $\mu$m images of a $67\arcsec$x$67\arcsec$
region centered on $\eta$ Corvi are shown in Fig.~\ref{fig:scuba}.
Each $0\farcs 2$ pixel in the images represents the weighted mean of
all bolometer measurements falling
(a) within $6\arcsec$ at 850 $\mu$m,
(c) within $7\arcsec$ at 450 $\mu$m,
and (e) within $4\arcsec$ at 450 $\mu$m.
The error is not constant across the image because of
the different noise levels in different bolometers, but its distribution in
the region plotted can be approximated by Gaussians with
(a) $5.1 \pm 0.8$, (c) $19 \pm 1.4$,
and (e) $34 \pm 3$ $\mu$Jy arcsec$^{-2}$.
This uncertainty is quantified in
Figs.~\ref{fig:scuba}b, \ref{fig:scuba}d, and \ref{fig:scuba}f
which show the corresponding signal to noise maps.
The same rebin methods were applied to the calibration data which showed
that point sources rebinned in this way are slightly smoothed and can be
fitted by Gaussians of FWHM
(a) 15\farcs 8, (c) 13\farcs 7, and (e) 9\farcs 5.

Extended emission was detected centered on the star (hereafter source A)
at each wavelength with greater than $5 \sigma$ confidence.
Another source (hereafter source B) was also detected with
$>4 \sigma$ confidence at both wavelengths at $\sim 24\arcsec$ NNW of
$\eta$ Corvi.
The $100\arcsec$x$100\arcsec$ field around the star was analyzed
to determine the noise level.
Structure at $<3 \sigma$ confidence is found randomly distributed across
the field of view and has the same level and distribution as that
expected from noise.\footnote{The expected noise distribution was
determined by simulating an observation of a blank field:
each bolometer was assumed to see noise with a Gaussian distribution
with the same standard deviation as the observed scatter in the
measurements for that bolometer (in that observation);
these data were then rebinned in the same way as the observations in
Fig.~\ref{fig:scuba}.}
Thus the zero levels in all maps were set to show only structure
detected with $>3\sigma$ confidence.
The remaining field is blank on this scale apart from
a source (hereafter source C) detected with $>3\sigma$ confidence
$50\arcsec$ SW of $\eta$ Corvi at 850 $\mu$m but not at 450 $\mu$m;
the edge of source C can just be seen in the bottom left corner of
Figs.~\ref{fig:scuba}a and \ref{fig:scuba}b.
The fluxes, locations and morphologies of these sources are discussed in
more detail below.

\subsubsection{Source A: Circumstellar disk}
\label{ss:srca}
The extended emission found centred on the star (source A) we attribute
to a circumstellar disk detected by IRAS in the far-IR
(e.g., Stencel \& Backman 1991).
Fitting a Gaussian to the rebinned 850 $\mu$m data
indicates that the 850 $\mu$m emission has a FWHM of $19\farcs 4 \pm
0\farcs 2$ and is centered on the star to within $\pm 2\arcsec$.
Subtracting the observed PSF (which has a FWHM of $15\farcs 8$)
in quadrature from this measurement implies that the disk is significantly
extended on a scale of $11\farcs 3 \pm 0\farcs 4$, indicative of a disk of
100 AU radius.
There is no evidence for any azimuthal asymmetry in the 850 $\mu$m
emission.

The structure of the disk is also evident in the 450 $\mu$m observation
at much higher resolution.
The 450 $\mu$m data are presented in two ways:
Figs.~\ref{fig:scuba}c and \ref{fig:scuba}d show the data rebinned to
a resolution of $13\farcs 7$,
while Figs.~\ref{fig:scuba}e and \ref{fig:scuba}f were
rebinned to a resolution of $9\farcs 5$.
The former images are appropriate for studying the disk's larger scale
structure, while the latter images show its high resolution structure.

The highest contours in Figs.~\ref{fig:scuba}c and
\ref{fig:scuba}d show that the emission is elongated along
a position angle of $130 \pm 10^\circ$.
The origin of this axis of symmetry is seen more clearly in
the higher resolution image of Fig.~\ref{fig:scuba}e
in which the emission is resolved into two peaks. 
These peaks are at an RA and Dec offset of $-1\farcs 8$, $-3\farcs 7$
(at 0.15 mJy arcsec$^{-2}$) and $+7\farcs 1$, $2\farcs 8$
(at 0.13 mJy arcsec$^{-2}$).
Given that the two peaks are detected at the $4.9\sigma$ and
$4.5\sigma$ levels, the location of each peak is not known with
better precision than $\pm 1\farcs 8$ (beam size divided by
signal-to-noise), and the peaks are equally bright to within the
uncertainties.
\footnote{The uncertainties in clump location and flux were confirmed
by analyzing the effect of noise on models which, in the absence of
noise, reproduce two equidistant peaks of equal brightness (\S
\ref{ss:ring}).}
Thus the observed structure is entirely consistent with two equally
bright peaks at 0.14 mJy arcsec$^{-2}$ that are equidistant from the
star at an offset of $5\farcs 5$ (100 AU) and a position angle of $130
\pm 10^\circ$ (counterclockwise from north).
Since this is the simplest interpretation of the data, it is the one we
adopt in the rest of the paper.

The total flux in a $30\arcsec$ ($40\arcsec$) aperture centered on
the star is $14.3 \pm 1.3$ ($17.1 \pm 1.8$) mJy at 850 $\mu$m
and $58.2 \pm 6.2$ ($74.9 \pm 8.5$) mJy at 450 $\mu$m.
Adding the calibration error in quadrature gives the uncertainty of
these measurements as $\pm 1.8$ (2.4) mJy at 850 $\mu$m and
$\pm 9.8$ (12.9) mJy at 450 $\mu$m.
These fluxes imply disk emission with a spectrum
$\propto \lambda^{-2.2 \pm 0.5}$ in the sub-mm, consistent with
black body emission in the Rayleigh-Jeans regime.
The 850 $\mu$m flux is almost a factor of two higher than that
presented by Sheret et al. (2004) who measured
$7.5 \pm 1.2$ mJy for this source.
However, the two datasets are consistent because their
observations were performed in photometry mode which is
insensitive to emission extended on scales larger than the beamsize;
Sheret et al. noted that the fluxes they derived could
underestimate the true flux by a factor of 2 if the source
is extended.

\subsubsection{Sources B and C}
\label{ss:srcbc}
Source B is located $8\arcsec$ W and $25\arcsec$ N of $\eta$ Corvi at 850
$\mu$m.
Its measured FWHM of 14\arcsec at 850 $\mu$m is not significantly different
to that of a point source, and its total flux is $6.2 \pm 0.8$ mJy in an
aperture of $18\arcsec$ diameter\footnote{The width of this aperture was
chosen to maximise the signal to noise of the detection while not losing
the flux at large offsets from the center.}.
This source thus provides an independent measure
of the PSF in the image for comparison with source A.
At 450 $\mu$m the emission peaks at $9\arcsec$ W and $21\arcsec$ N of
$\eta$ Corvi and has a total flux of $19.0 \pm 4.1$ mJy in
an $18\arcsec$ aperture.
Source C was discovered in the 850 $\mu$m map at an offset of
$26\arcsec$ E, $39\arcsec$ S of $\eta$ Corvi with a total flux of 
$3.9 \pm 0.8$ ($5.4 \pm 1.3$) mJy in an $18\arcsec$ ($30\arcsec$) aperture.
No emission was detected from the location of source C in the 450 $\mu$m
image at the $3\sigma$ upper limit level of $<12$ (21) mJy in an
$18\arcsec$ ($30\arcsec$) aperture.
This implies a spectrum of emission $\propto \lambda^{>-1.8}$.

\subsubsection{Background Confusion}
\label{ss:back}
Several background sources are expected in any given sub-mm image.
Since this star is not in the Galactic plane (it is at a Galactic
latitude of $46^\circ$) and is far from known regions of star
formation, the density of background sources in this image would be
expected to be the same as that measured in similar regions.
Scott et al. (2002) estimate that there are about 620 sources
per square degree above 5 mJy at 850 $\mu$m, while there are
as many as 2000 sources per square degree above 3 mJy at
850 $\mu$m (Eales et al. 2000).
Smail et al. (2002) estimate that there are some 2000 $>10$ mJy
sources at 450 $\mu$m per square degree.
Thus in the field of view analyzed in this paper,
$100\arcsec$x$100\arcsec$,
we would expect to find: 1.5 sources above 3 mJy in our 850 $\mu$m
image and 1.5 sources above 10 mJy in our 450 $\mu$m image.
These number counts are consistent with sources B and C
being background objects and from now on we interpret them as such.
The probability of two background sources (at
either $>3$ mJy or $>10$ mJy at 850 and 450 $\mu$m respectively)
falling within $10\arcsec$ of $\eta$ Corvi, and so providing confusion
with source A, is less than 1:300;
the chance of one such object falling within this distance is 1:16,
so background objects should be considered as a potential source of
uncertainty in the low level morphology of this source.

While source B is interpreted as a background object, we
note that a 19 mJy source was observed at an offset of
$34\arcsec$ (660 AU projected separation) in 850 $\mu$m images
of $\beta$ Pictoris (Holland et al. 1998).
Thus it is possible that offset sub-mm clumps may be more
commonly associated with debris disks.
The spectral slope of source B, emission $\propto
\lambda^{-1.8 \pm 0.5}$, is similar to that of source A, meaning
that it is not inconsistent with the sources being comprised
of dust with similar properties.
However, this spectral slope is also similar to that of other
background sub-mm objects ($F_{450}/F_{850} = 3.4 \pm 0.6$,
Smail et al. 2002) and further study is required to
determine the true nature of this object.
If associated with the star the projected separation of source B
(480 AU) would make it a highly unusual object.

\subsection{Mid-IR}
\label{ss:midirresults}
\begin{figure}
  \begin{center}
    \begin{tabular}{c}
        \epsscale{0.9} \plotone{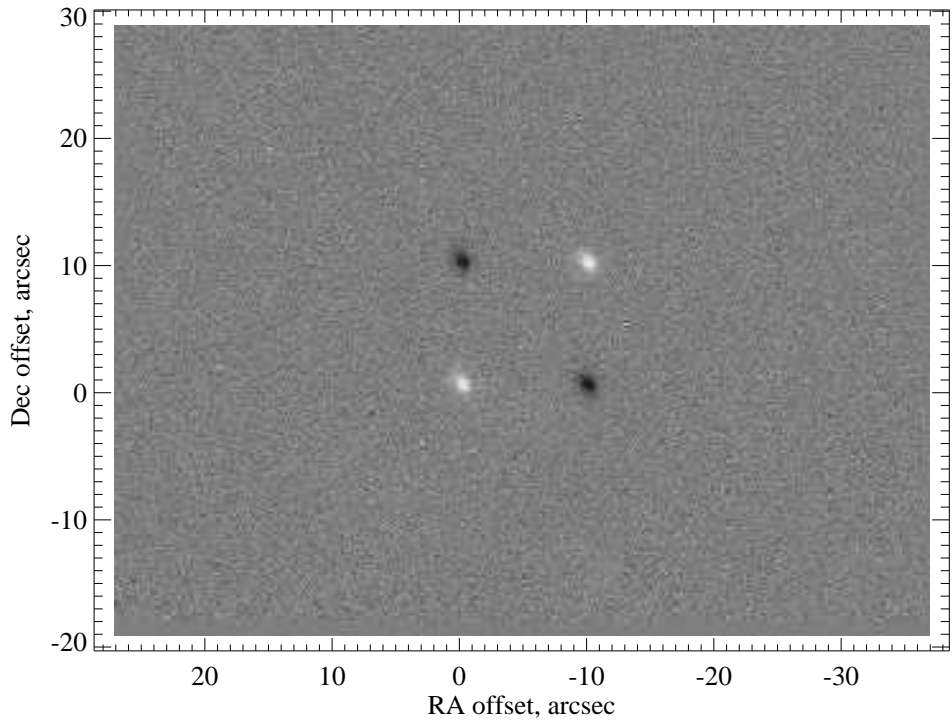}
    \end{tabular}
  \end{center}
  \caption{TIMMI2 observation of $\eta$ Corvi at 11.9 $\mu$m.
  The greyscale runs linearly from -40 to 40 mJy/pixel.
  The pattern of two positive and two negative sources is caused
  by the observing strategy and is that expected from
  emission from just one source centered on the star.
  The remainder of the image has zero mean with a pixel-to-pixel
  deviation which is flat across the image at
  $1\sigma = 3.6$ mJy/pixel.
  \label{fig:midir}}
\end{figure}

The image resulting from the 11.9 $\mu$m observations of $\eta$ Corvi
described in \S \ref{ss:midir} is presented in Fig.~\ref{fig:midir}.
The pattern of positive and negative sources shown in this image
is that expected from a single source centered on the star.
Each of the peaks was detected with a statistical signal to noise of $>23$,
and is consistent with point source emission (FWHM of $1\farcs 21 \pm
0\farcs 02$).
The total measured flux was $1.51 \pm 0.24$ Jy ($SNR=7$);
the dominant uncertainty in this flux estimate is the calibration
uncertainty. 
This is consistent, within the error bars of the measurement, with
the $1.642 \pm 0.038$ Jy at 12 $\mu$m measured by IRAS (Table \ref{tab:fluxes}).
Further point-like sources within the field of view, including at
the location of source B, were ruled out at $<70$ mJy
with $3\sigma$ certainty.

\section{Modeling}
\label{s:disc}

\subsection{Modeling the 450 $\mu$m images}
\label{ss:ring}
A disk structure characterized by two bright lobes equidistant from
the star has previously been seen in the disks of Fomalhaut and
HR4796A and interpreted as evidence of an edge-on dust ring (e.g.,
Holland et al. 1998; Telesco et al. 2000; Holland et al. 2003).
However, two clumps straddling the star have also been
observed and interpreted as azimuthal asymmetries in a face-on
dust ring around Vega (Holland et al. 1998; Koerner et al. 2001;
Wilner et al. 2002).
Vega's disk was assumed to be face-on due to the symmetric nature
of the lower level contours (Holland et al. 1998).
In this respect the $\eta$ Corvi disk bears more resemblance to
Vega than to Fomalhaut:
In Fig.~\ref{fig:scuba}c the source is significantly resolved at
all position angles, and is only slightly elongated with a FWHM
of $24\arcsec$ along a position angle of $130^\circ$ and $22\arcsec$
at $40^\circ$.
Subtracting these observed widths in quadrature from the FWHM of a
point source ($13\farcs 7$), we infer that this emission comes from
a ring at moderate inclination ($\sim 60^\circ$ from edge-on) to our
line-of-sight.

We simulated the 450 $\mu$m image by making a model of the predicted
emission distribution, smoothing it by the PSF ($8\farcs 1$ Gaussian),
then using the resulting image to determine the expected flux measured
by each bolometer in each jiggle observation in the SCUBA data and
rebinning the model in the same way as the observation
(Fig.~\ref{fig:scuba}).
A noisy model was also made by adding Gaussian noise at the same
level as the deviation in the measurements for bolometers in individual
observations to determine the extent to which noise affects the observed
structure.

\begin{figure}
  \begin{center}
    \begin{tabular}{c}
        \epsscale{0.8} \plotone{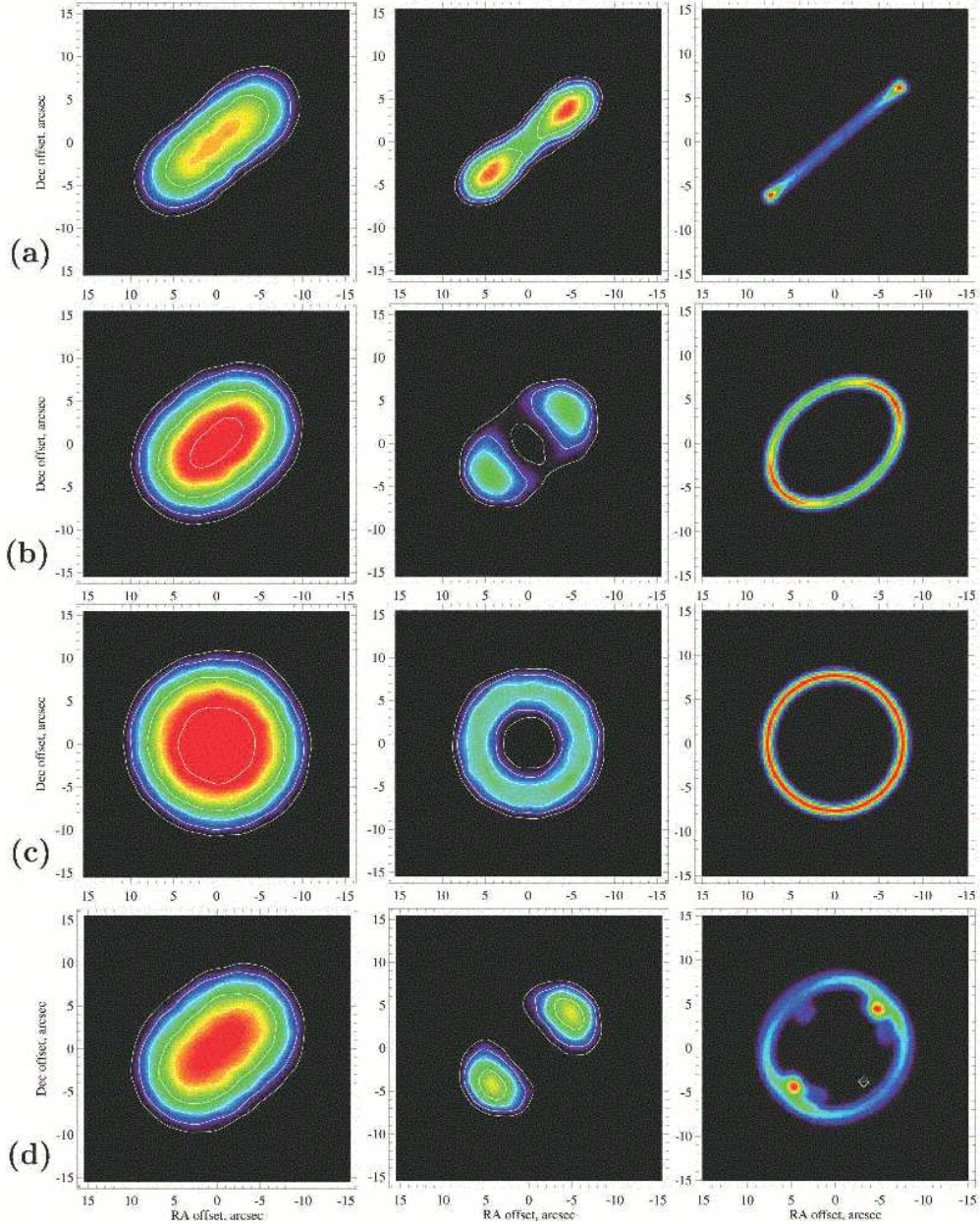}
    \end{tabular}
  \end{center}
  \caption{Models of the 450 $\mu$m disk structure:
  \textbf{(a)} 180 AU radius ring seen edge-on;
  \textbf{(b)} 150 AU ring inclined $45^\circ$ to our
  line-of-sight;
  \textbf{(c)} 140 AU ring seen face-on;
  \textbf{(d)} ring perturbed by the migration of a Neptune mass
  planet from 80-105 AU over 25 Myr seen inclined $60^\circ$
  (from edge-on) to our line-of-sight.
  The left hand panels show the simulated observations rebinned and
  plotted in the same way as Fig.~\ref{fig:scuba}c (i.e., with an
  effective resolution of $13\farcs 7$ and using the same color scale
  and contour levels).
  The middle panels are the equivalent plots for comparison with the
  rebin shown in Fig.~\ref{fig:scuba}e (i.e., with an effective
  resolution of $9\farcs 5$).
  The right hand panels show the structure of the disk models at
  $1\arcsec$ resolution.
  The location of the planet in model \textbf{(d)} is shown with a
  diamond;
  however, the same model rotated by $180^\circ$ provides an equally
  valid fit to the observation.
  \label{fig:mod}}
\end{figure}

The simplest model which was considered was one in which the emission comes from
an axisymmetric torus of width $dr = 10$ AU which is inclined to our line-of-sight.
The radius and surface density of the ring were constrained by the FWHM
and peak fluxes in the resulting images, leaving the
inclination of the ring as the only free parameter.
Simulated observations of models inclined by $0^\circ$, $45^\circ$,
and $90^\circ$ from edge-on are shown in 
Figs.~\ref{fig:mod}a, \ref{fig:mod}b, and \ref{fig:mod}c, respectively;
the radii of the rings in these models were determined to be 140,
150, and 180 AU respectively, all with errors of $\pm 10$ AU.
It is possible to rule out the edge-on model at the statistically
significant level, because this model does not fit the run of flux vs distance
from the peak along a position angle of $40^\circ$
(the model is unresolved, FWHM of $13\farcs 7$, in the left panel of
Fig.~\ref{fig:mod}a).
This means that while the peak fluxes in the images are well
reproduced, the total flux within 15\arcsec from this model is just
37 mJy, more than $3\sigma$ below that observed.
The face-on model is also ruled out at a statistically signficant
level because it fails to reproduce an axis of symmetry.
This symmetry can be shown to be significant, since the total flux
measured by bolometers falling within 10\arcsec of the star and within
$30^\circ$ of a position angle of $120^\circ$ is $17.2 \pm 2.1$ mJy,
whereas the same measurements for position angles centered on
$0^\circ$ and $60^\circ$ are $8.3 \pm 2.1$ and $5.6 \pm 2.0$ mJy
respectively.
We conclude that if this disk is axisymmetric, an inclination of
$\sim 45 \pm 25^\circ$ to our line-of-sight is most likely, and its
radius is $150 \pm 20$ AU.
An inclination of $45-60^\circ$ from edge-on is consistent with 
the disk being aligned with the stellar equator given that the star
$\eta$ Corvi has $v\sin{i} = 68$ km s$^{-1}$ (Mora et al. 2001) whereas
the mean $v\sin{i}$ for F0V stars is 106 km s$^{-1}$ (Abt \& Morrell
1995).

An inclination of $\geq 60^\circ$ is still possible, however, if,
like the Vega disk, the structure of the $\eta$ Corvi disk includes
two clumps, one on either side of the star.
Such an axis of symmetry is to be expected if the disk is being
perturbed by planets.
Under a wide variety of conditions two clumps approximately equidistant
from the star are expected, either because P-R drag makes the dust
migrate into the exterior mean motion resonances of a massive planet
(Kuchner \& Holman 2003) or because the parent planetesimals feeding
the dust disk were trapped into the resonances of a planet when that
planet migrated outward (Wyatt 2003).
In Fig.~\ref{fig:mod}d we show how the 450 $\mu$m disk
structure can be explained by a disk inclined by $60^\circ$ to our
line-of-sight in which a Neptune mass planet ($17M_\oplus$) migrated out from
80 to 105 AU over a period of 25 Myr and so trapping planetesimals
into its 3:2 and 4:3 resonances (Wyatt 2003).
In this model the planet is at an offset of $5\arcsec$ at a
position angle of either 40 or $220^\circ$.
The planet's mass (and the time it took to complete its migration)
are poorly constrained in this model because a wide variety of
parameters would cause two clumps (Wyatt 2003).

We have presented two possible models which fit the 450
$\mu$m observations to within the uncertainty due to noise
(Figs.~\ref{fig:mod}b and \ref{fig:mod}d).
Clearly further observations are required to determine the
true morphology and orientation of the disk.
High resolution observations, such as those provided by
sub-mm or mm interferometry or optical coronography, would
be able to distinguish between the two models presented here,
or tell if an alternative model is required.

Regardless of the axisymmetry of the dust ring and its
inclination, the 450 $\mu$m images show there is a lack of
sub-mm emitting dust inside 100 AU.
To understand this, consider that the models presented in
Figs.~\ref{fig:mod}b and \ref{fig:mod}d fit the surface brightness
distribution at both $>150$ AU from the star and at the stellar
location, even though there is no dust at $<100$ AU in the models.
The reason is that emission at the stellar location is seen from the
150 AU ring because of the resolution of the observation.
We set a conservative constraint on the emission from an unresolved
sub-mm component from the $5\sigma$ uncertainty in the surface
brightness at the center of Figs.~\ref{fig:scuba}c and \ref{fig:scuba}e;
i.e., an additional unresolved component is ruled out at $<19$ mJy
at 450 $\mu$m, or less than one third of the total disk flux.

\subsection{Modeling the SED}
\label{ss:sed}
\begin{figure}
  \begin{center}
    \begin{tabular}{c}
        \epsscale{0.9} \plotone{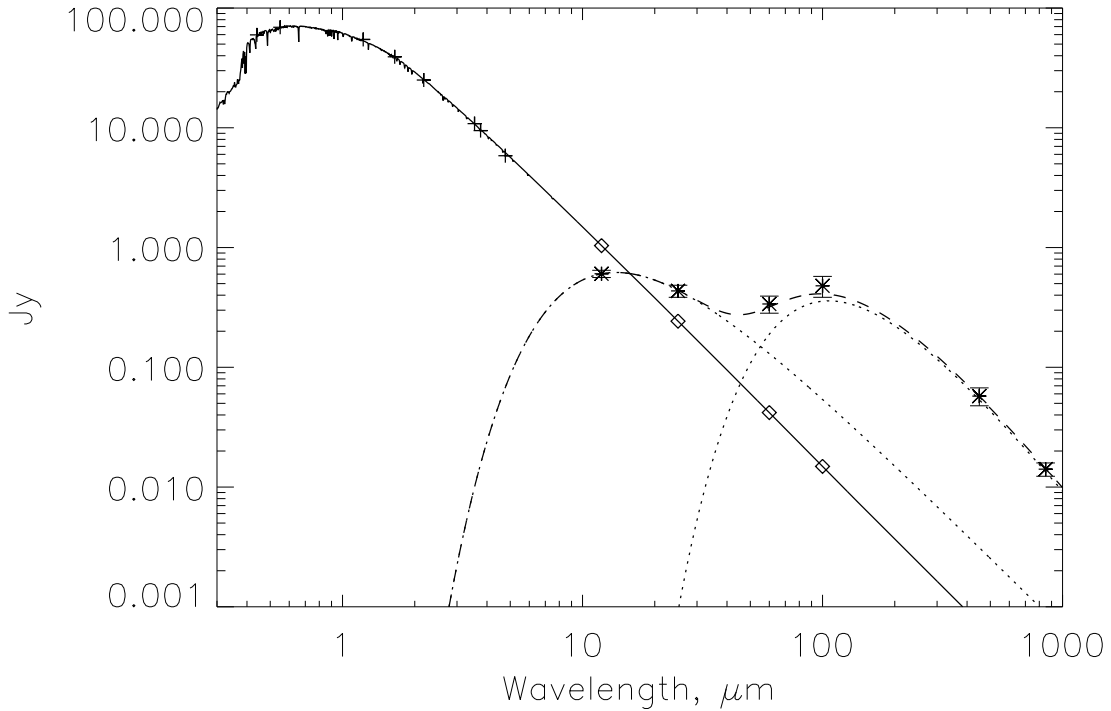}
    \end{tabular}
  \end{center}  
  \caption{Spectral energy distribution (SED) of emission from $\eta$ Corvi.
  The solid line shows the stellar spectrum modeled as a 7080 K Kurucz
  atmosphere and scaled to the near-IR fluxes in Sylvester et al.
  (1996) (pluses).
  The color corrected excess fluxes (i.e., after the estimated
  photospheric level, shown with diamonds, have been subtracted)
  are shown with asterisks and $1\sigma$ error bars.
  This excess emission has been modeled using a two component (hot and cool)
  model, each of which fitted by a modified black body;
  the hot component has a temperature of $370 \pm 60$ K, while the cool
  component is at $40 \pm 5$ K.
  The individual components are shown with dotted lines and the total flux
  with dashed line.
  \label{fig:sed}}
\end{figure}

The SED of $\eta$ Corvi from optical to sub-mm wavelengths
is plotted in Fig.~\ref{fig:sed}.
The IRAS fluxes were determined from SCANPI (IRAS Scan Processing and
Integration), available at the IRSA website (http://irsa.ipac.caltech.edu)
and the new values (including a previously unpublished detection at
100 $\mu$m) are given in Table 1.
The stellar fluxes at mid-IR to sub-mm wavelengths were determined from
a 7080 K Kurucz atmosphere scaled to the near-IR fluxes in Sylvester et
al. (1996).
Color corrected stellar fluxes were subtracted from the
IRAS fluxes to give the excess fluxes (which were also
color corrected).
Excess emission is found at all wavelengths from 12-850 $\mu$m.
As noted by previous authors, the spectrum of this excess emission is
such that it cannot all be accounted for by dust at a single
temperature or distance from the star (Sheret et al. 2004).

Mid-IR spectroscopy presented in Sylvester \& Mannings (2000)
found that most of the 12 $\mu$m excess emission must be centered
on the star.
It is also important that our 11.9 $\mu$m observations did not detect
any additional nearby sources which could be contributing to the 12
$\mu$m flux measured by IRAS (which could originate anywhere within
30\arcsec of the star; Lisse et al. 2002);
in particular source B does not have a significant counterpart at 12
$\mu$m.
Combining these findings implies that most of the IRAS emission is
centered on a region around the star that is smaller than our measured
FWHM, i.e., at $<10$ AU. 
Thus we modeled the disk as a single temperature cool component (causing
the excess at $\lambda \geq 60 \mu$m) with an additional single temperature
hot component (causing the excess at $\lambda \leq 60 \mu$m).
The fluxes from the individual components were modeled assuming
a black body emission distribution modified by a ratio
$(\lambda_0/\lambda)^\beta$ at $\lambda > \lambda_0$
(e.g., Dent et al. 2000).

The cool component is well fit by emission at $T=40 \pm 5$ K,
$\beta=0.5$ and $\lambda_0=20$ $\mu$m.
The fractional luminosity of this component is
$f=L_{ir}/L_\star=3 \times 10^{-5}$ and the inferred dust
mass assuming an opacity of $\kappa_{850} = 0.17$ m$^2$ kg$^{-1}$
(e.g., Wyatt, Dent \& Greaves 2003 and references therein) is
$0.04 M_\earth$.
The black body temperature at a distance of 150 AU is 34 K.
Thus the dust is only slightly warmer than black body, implying
that the dust is comprised of grains larger than a few 10s of $\mu$m.
This is consistent with the inferred $\beta$ which is comparable
with that of debris disks of similar age (Dent et al. 2000).
In a further study we performed a more detailed SED model using realistic
grain properties (see, e.g., Wyatt \& Dent 2002; Sheret et al. 2004)
and found that this spectrum could be fitted with a dust ring at 150 AU
with the same dust size distribution as that expected from a
collisional cascade, but one with an additional imposed cut-off for grains
smaller than $D_{min} = 30$ $\mu$m rather than at $\sim 4 \mu$m as expected
from radiation pressure blow-out (Wyatt et al. 1999).
Grains in the size range of 4-30 $\mu$m must be absent
because they are too hot ($>40$ K) at this distance from the star to
explain the shape of the SED, which would have stronger emission
at 25 and 60 $\mu$m if these grains were present.
Possible reasons for the absence of small grains,
also seen in the spectrum of $\epsilon$ Eridani, were discussed
in Sheret et al. (2004).
These include the possibility that the $4-30$ $\mu$m grains in
the outer disk of $\eta$ Corvi are destroyed in collisions with $<4$
$\mu$m grains which are in the process of being blown out of
the inner regions by radiation pressure (Krivov et al. 2000).
We also note that a dearth of small grains is expected in the
model in which clumpy structure is formed by planet migration
(Wyatt 2003).

The fit presented in Fig.~\ref{fig:sed} shows the hot component has
a temperature of $370 \pm 60$ K (assumed to be black body, i.e.,
$\beta=0$) and has a fractional luminosity of $f=5 \times 10^{-4}$.
While this fractional luminosity is an order of magnitude higher than
that of the cool component, the mass of the hot component is inferred
to be at least two orders of magnitude less than that of the cold
component (because of its higher temperature and lower inferred
sub-mm flux).
This temperature implies that the dust lies 1-2 AU from the star.
While this is the interpretation discussed in the rest of this paper,
we note that this distance should be regarded as suggestive rather
than definitive, since the two temperature model used to fit the
SED may be an oversimplification of the true dust distribution.
It is also possible that the dust lies in a more extended distribution
wherein the 12 $\mu$m excess comes from dust which is hotter than 370 K
(and so is unresolved), and the 25 $\mu$m excess comes from cooler dust
located between 1 and 150 AU.
High resolution mid-IR observations are required to determine the
location of the 25 $\mu$m emission.

\section{Discussion}
\label{s:concl}
The age of this star has been estimated in the literature to 
be 1.3 Gyr from its position relative to evolutionary tracks in the HR
diagram (Mallik, Parthasarathy \& Pati 2003).
Such a high age is also consistent with the low X-ray
luminosity of $\eta$ Corvi.
H\"{u}nsch et al. (1999) found this star to have
$L_X = 47.8 \times 10^{27}$ erg s$^{-1}$, a value which is
significantly below that of members of the Pleiades
with similar $B-V=0.4$ (Micela et al. 1999), implying an age
of $>125$ Myr.
On the other hand, this X-ray luminosity lies very close to the mean
value of A-F stars in the Hyades cluster (Stern, Schmitt \& Kahabka 1995)
implying an age of 600-800 Myr.
Here we adopt an age of 1 Gyr for $\eta$ Corvi.

Given such a large age, the dust in this system cannot
be primordial, since this is much longer than the P-R drag lifetime of
the dust: dust with $F_{rad}/F_{grav} = 0.5$ would migrate from 150 AU
to 0 AU (where it would evaporate) on timescales of 20 Myr
(Burns et al. 1979).
This is unlikely to be the ultimate fate of the dust originating
at 150 AU, however, since the collisional lifetime of this dust is
$\sim 1$ Myr, so it is more likely to be destroyed in mutual
collisions (and then removed by radiation pressure) before it
reaches the inner regions.
In any case, the dust must be replenished, presumably by the
collisional destruction of a population of planetesimals orbiting
at 150 AU from $\eta$ Corvi.
Using the same collisional modeling as that presented in Wyatt \& Dent
(2002), we infer from the age of 1 Gyr that the collisional cascade
in this system starts with planetesimals a few km in size implying
a total mass of $20 M_\oplus$ in planetesimals in the ring at 150 AU.
\footnote{In this model, the collisional lifetimes of planetesimals
of different sizes are computed using a model for the outcome
of different collisions and assuming a size distribution appropriate
for material in a collisional cascade scaled to the surface area
of dust in that distribution through a fit to the SED of the disk
emission.}
This is similar to the mass inferred for the collisional cascade of
the disk around star Fomalhaut which has an age of 200 Myr
(Wyatt \& Dent 2002).

The relative lack of dust at $<100$ AU indicates that there is a significant
lack of colliding planetesimals in this region.
This could be because planet formation has proceeded to such a
stage that the planetesimals have grown into planets (e.g., Kenyon
\& Bromley 2002);
the perturbations from the outermost planet of that system are
potentially observable in the structure of the outer
ring (\S \ref{ss:ring}).
The dust at a few AU must also be replenished.
Just like the zodiacal cloud in the solar system, this replenishment
could come from either the collisional destruction of planetesimals
in an asteroid belt at a few AU or in the sublimation of comets,
possibly comets which originated in the cooler outer ring.
Mid-IR spectroscopy may be able to shed light on the composition
of this dust, and so be able to determine whether this dust formed
in a hot region close to the star or if it formed in a cold region
far from the star.

The age of 1 Gyr means that it is very surprising that
this star has detectable hot emission.
The only other stars with mid-IR excesses are young;
e.g., Laureijs et al. (2002) found no stars older than 400 Myr
with excess 25 $\mu$m emission and
Weinberger et al. (2004) found that even at 10 Myr, very few
stars have significant mid-IR excess.
While it is possible that this is a system which has always
had an unusually massive hot dust population, another
possibility is that the mid-IR excess is a transient phenomenon.
An increase in mid-IR excess might be expected after the
break-up of a massive asteroid in an asteroid belt at a few AU
(Grogan, Dermott, \& Durda 2001), or after an event which perturbed
comets from the ring at 150 AU into the inner region.
Such an event could have been the recent close passage of a
nearby star (Larwood \& Kalas 2001), or the recent scattering
and subsequent migration of a Neptune-sized planet from its
formation at a few 10s of AU out to $\sim 100$ AU (e.g., Thommes,
Duncan \& Levison 2002).

\acknowledgments
The JCMT is operated by the Joint Astronomy Centre on behalf
of the UK Particle Physics and Astronomy Research Council,
the Canadian National Research Council and the Netherlands Organization
for Scientific Research.
This research has made use of the NASA/ IPAC Infrared Science
Archive, which is operated by the Jet Propulsion Laboratory,
California Institute of Technology, under contract with the
National Aeronautics and Space Administration.
This research was supported in part by PPARC funding.
The careful analysis of the referee, C. M. Lisse, has also
greatly improved this paper.


\clearpage

\clearpage

\clearpage

\clearpage

\clearpage

\begin{deluxetable}{cccccc}
  \tabletypesize{\scriptsize}
  \tablecaption{Flux of emission from $\eta$ Corvi in mJy:
  top --- total measured flux;
  middle --- stellar flux extrapolated from near-IR fluxes presented in
  Sylvester et al. (1996);
  bottom --- excess flux.
  Fluxes at 12-100 $\mu$m have been determined from a reanalysis of the IRAS
  scans using SCANPI.
  Excess IRAS fluxes have been determined by subtracting a color corrected
  stellar flux from the total flux, then color correcting this excess assuming
  the spectrum is flat ($F_\nu \propto \nu^0$; see
  Fig.~\ref{fig:sed}).
  These color corrections are the reason the stellar and excess fluxes
  do not add up to the total measured flux.
  Sub-mm fluxes are the total flux within $30\arcsec$ of the star
  from the SCUBA observations presented in this paper.
  \label{tab:fluxes}}
  \tablewidth{0pt}
  \tablehead{\colhead{$F_{12}$} & \colhead{$F_{25}$} &
    \colhead{$F_{60}$} & \colhead{$F_{100}$} & \colhead{$F_{450}$} & \colhead{$F_{850}$}}
  \startdata
    $2170 \pm 42$ & $820 \pm 50$ & $410 \pm 55$ & $500 \pm 95$ & $58.2 \pm 9.8$  & $14.3 \pm 1.8$    \\
    1039          & 243          & 42           & 15           & 0.7             & 0.2               \\
    $603 \pm 38$  & $434 \pm 46$ & $338 \pm 52$ & $479 \pm 94$ & $57.5 \pm 9.8$  & $14.1 \pm 1.8$    \\
  \enddata
\end{deluxetable}

\end{document}